\pdfoutput=1

\documentclass[11pt]{article}

\usepackage[final]{acl}

\usepackage{times}
\usepackage{latexsym}

\usepackage[T1]{fontenc}

\usepackage[utf8]{inputenc}

\usepackage{microtype}

\usepackage{inconsolata}

\usepackage{graphicx}

\usepackage{babel}
\usepackage{multirow}
\usepackage{tikz}
\usepackage{float}

\usepackage{amsmath}
\usepackage{booktabs}
\usepackage{pgfplots}
\usepackage{xcolor,colortbl}
\usepackage{amsfonts}

\pgfplotsset{compat=1.17}

\usetikzlibrary{positioning, arrows.meta, calc}
\usetikzlibrary{bayesnet}
\usetikzlibrary{arrows}

\newcommand{\todo}[1]{}
\renewcommand{\todo}[1]{{\color{red} TODO: {#1}}}

\title{Improving Automatic Speech Recognition with Decoder-Centric Regularisation in Encoder-Decoder Models}

\author{Alexander Polok, Santosh Kesiraju, Karel Beneš, Lukáš Burget, Jan Černocký
  \\[1ex]
  Speech@FIT, Brno University of Technology \\[1ex]
  \href{mailto:ipoloka@fit.vut.cz}{ipoloka@fit.vut.cz}
}

\begin{document}
\maketitle
\begin{abstract}
This paper proposes a simple yet effective way of regularising the encoder-decoder-based automatic speech recognition (ASR) models that enhance the robustness of the model and improve the generalisation to out-of-domain scenarios.
The proposed approach is dubbed as $\textbf{De}$coder-$\textbf{C}$entric $\textbf{R}$egularisation in $\textbf{E}$ncoder-$\textbf{D}$ecoder (DeCRED) architecture for ASR, where auxiliary classifier(s) is introduced in layers of the decoder module. Leveraging these classifiers, we propose two decoding strategies that re-estimate the next token probabilities. 
Using the recent E-branchformer architecture, we build strong ASR systems that 
obtained competitive WERs as compared to Whisper-medium and outperformed OWSM v3; while relying only on a fraction of training data and model size. On top of such a strong baseline, we show that DeCRED can further improve the results and, moreover, generalise much better to out-of-domain scenarios, where we show an absolute reduction of 2.7 and 2.9 WERs on AMI and Gigaspeech datasets, respectively. We provide extensive analysis and accompanying experiments that support the benefits of the proposed regularisation scheme.  
\end{abstract}

\section{Introduction}
One of the key challenges in automatic speech recognition (ASR) is the ability of the models to generalise to new or unseen domains.
Large-scale training on multiple domains~\citep{narayanan_toward_2018}, data augmentation, multi-task training~\cite{hori_joint_2017}, architecture-specific regularisation~\citep{lee_intermediate_2021} are some of the strategies for improving the robustness of ASR systems. Some of these techniques, such as SpecAug~\citep{park_specaugment_2019}, joint CTC/attention training, label smoothing~\cite{PereyraTCKH17,kim18f_interspeech} are now a defacto and are integrated into open source toolkits~\cite{watanabe_espnet_2018,speechbrain}. Recent years have seen a shift towards large-scale training~\cite{chan_speechstew_2021} of speech models such as Whisper from OpenAI~\citep{radford_robust_2023}.
Despite its impressive recognition accuracy on many research datasets, the lack of transparency about the training data has led the scientific community to build an \textit{open-source equivalent} of Whisper.
One such effort, dubbed as OWSM (Open Whisper-style Speech Model)~\cite{peng_reproducing_2023} is trained on publicly available speech datasets, using an open source toolkit ESPnet~\citep{watanabe_espnet_2018}. It is important to note that these datasets come from various domains and styles, such as conversational, lectures/talks, broadcast news, telephone-speech, read-speech, etc. The size of each dataset in the training set also varies significantly. However, it is hard to evaluate the out-of-domain generalization of these models since all the standard datasets were already seen during the training. 

In this paper, we ask the question \textit{what additional, yet, simple method can further improve the robustness of ASR systems}? To answer this, we create a setup where we train a large-scale ASR model\footnote{To the extent supported by the computational budget available to us.} on a collection of multiple datasets, carefully leaving out a few datasets to be used for out-of-domain evaluation. The ASR model is built on the most recent E-branchformer architecture~\cite{kim_e-branchformer_2023} and trained with all the aforementioned augmentation, label-smoothing and multi-task training techniques that are known the improve the robustness of the model. On top of that,
we hypothesise that regularising the ASR model during training prevents overfitting and helps generalise better in out-of-domain scenarios. Our choice of regularisation is architecture-driven, i.e., we choose to regularise the decoder module of the encoder-decoder architecture by introducing auxiliary classifier(s) in the intermediate layers. The decoder module in a standard encoder-decoder-based ASR can be viewed as an auto-regressive internal language model (ILM)~\cite{zeineldeen_investigating_2021}\footnote{Encoder with a CTC objective can also learn an internal language model, though non-autoregressive.}. Having an auxiliary classifier adds negligible computational cost during training and no additional cost during decoding (inference) if the auxiliary classifiers are ignored. We further hypothesise that these auxiliary classifiers can be exploited for rapid adaptation to a new domain that was not seen during training. 

\subsection{Summary and contributions}
\begin{itemize}
    \item Section~\ref{sec:related} discusses related works and highlights how our work complements the existing body of research.
    \item The decoder-centric regularisation is formally introduced in Section~\ref{sec:decred}, where we also describe the proposed decoding strategies that exploit the auxiliary classifiers for joint-decoding and rapid domain adaptation.
    \item Experiment protocol is described in Section~\ref{sec:exp}, followed by experiments on large-scale multi-domain training and evaluation on out-of-domain datasets; detail of which are presented in Section~\ref{sec:multi-domain}. Compared to the baseline, we show an absolute reduction of 2.7 and 2.9 word error rates (WERs) on AMI and Gigaspeech out-of-domain datasets, respectively.
    \item The analysis of the internal language model (ILM) for both the baseline and the proposed DeCRED is presented in Section~\ref{sec:ilm}. This analysis provides complementary evidence supporting our hypothesis that regularising the decoder module indeed helps generalise in out-of-domain scenarios.
    \item Experiments and results from an extensive ablation study are presented in Section~\ref{sec:ablation}, where we identify the various factors that affect the final performance (WER) of the systems.
    \item Finally, our implementations\footnote{
    \url{https://github.com/BUTSpeechFIT/DeCRED}
    } are built on top of open-source \texttt{transformers} library~\cite{wolf_transformers_2020}, facilitating easy replication of our results. We intend to release all model checkpoints along with the corresponding test hypotheses. Our code also allows for single-line inference within the HuggingFace ecosystem.
\end{itemize}

\section{Related works}
\label{sec:related}
The idea of auxiliary classifiers or intermediate regularizers has been explored in ASR and self-supervised learning models for speech representation. Most of the works use the auxiliary classifiers in the encoder module. For instance, \citet{lee_intermediate_2021} uses intermediate CTC objectives in the encoder module for ASR, while \citet{Nozaki_self_cond_2021} extends this by adding intermediate classifier outputs to the input of the next layer, conditioning the final layer's predictions on these intermediate outputs. Similarly, \citet{wang2021selfsupervised} employs a similar scheme for training self-supervised speech encoders. \citet{zhang2022intermediatelayer} regularizes both the encoder and decoder modules by passing the intermediate representations from the encoder directly to the intermediate layers in the decoder. While these works have demonstrated improvements over their respective baselines, our work complements the prior work in the following ways:
\begin{itemize}
    \item We introduce auxiliary classifier(s) only in the decoder module of the encoder-decoder architecture, essentially regularising the auto-gressive internal language model.
    \item We study the effect of such a regularisation scheme in the context of out-of-domain generalisation.
    \item We propose to exploit the auxiliary classifiers for rapid domain adaptation.
\end{itemize}

In the case of large-scale training of end-to-end ASR models, we mainly take inspiration from prior works such as SpeechStew~\citep{chan_speechstew_2021} and OWSM~\citep{peng_reproducing_2023}, where we simply mix multiple publicly available datasets to train our models. It is important to note that simple aggregation from multiple sources (datasets) without text normalising can cause the models to \textit{memorise} dataset-specific annotation styles~\cite{peng_reproducing_2023}; which is not desired for a general purpose ASR system. This also indicates a potential inefficiency, wherein model parameters are allocated towards recognising data sources rather than solving the intended task(s). As it is inevitable, we investigate and quantify the effect of text normalisation on the model's recognition performance.

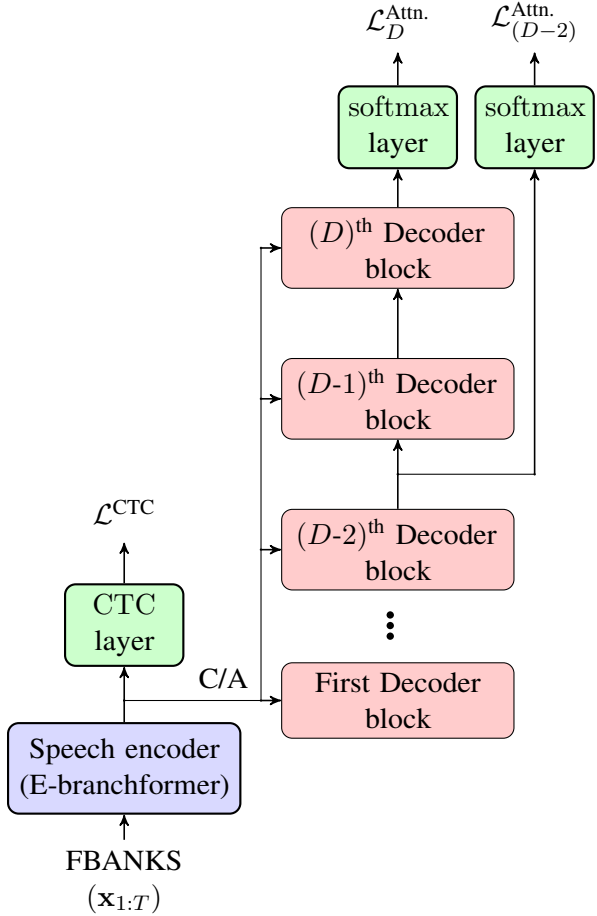
\begin{figure}[!t]
\centering
\begin{tikzpicture}
[
rect/.style={minimum size=0.2cm,minimum height=1.2cm,text width=28mm,
	align=center, rectangle,draw,rounded corners,thick, fill=blue!15},
rect2/.style={rectangle,minimum size=10mm,text width=28mm,   
    align=center,draw,rounded corners,fill=red!20},
emb/.style={minimum size=0.2cm,minimum height=0.5cm,text width=13mm,
    align=center, rectangle,draw,rounded corners,thick, fill=green!20},
shaded/.style={minimum size=0.1cm, minimum height=6mm, text width=0.5mm, align=center,   
	rectangle,draw,thick,fill=lightgray!50},
var/.style={text width=2.4cm,minimum height=8mm,rectangle,align=center},
eqn/.style={text width=8.5cm,minimum height=8mm,rectangle,align=center},
arr/.style={->,>=stealth',semithick},
outer/.style={draw, thick, densely dashed, fill=white!5,
	inner xsep=1ex, inner ysep=0ex, yshift=1ex,	fit=#1},
hidden/.style={circle,scale=0.05,minimum size=1pt,draw},
]
\node (x)  [minimum size=3cm,var]  at (-0.1, -0.1)  {FBANKS $(\mathbf{x}_{1:T})$};
\node (enc) [rect]   at (-0.1, 1.4) {Speech encoder \\ (E-branchformer)};
\node (ctc) [emb]    at (-0.1, 3.3) {$\mathrm{CTC}$ \\ layer};
\node (pctc) [var]   at (-0.1, 4.8) {$\mathcal{L}^{\text{CTC}}$};

\node (dec0) [rect2] at (3.5, 2.3) {First Decoder \\ block};
\node (dt)  [var]    at (3.4, 3.4)   {\Huge $\boldsymbol{\vdots}$};
\node (dec1) [rect2] at (3.5, 4.3) {$(D\text{-}2)^{\textrm{th}}$ Decoder \\ block};
\node (h13) [hidden] at (3.5, 5.3)   {};
\node (sf1) [emb]    at (5.3, 9.9) {$\mathrm{softmax}$ \\ layer};
\node (pd1) [var]    at (5.3, 11.3) {$\mathcal{L}_{(D-2)}^{\text{Attn.}}$};
\node (dec2) [rect2] at (3.5, 6.3) {$(D\text{-}1)^{\textrm{th}}$ Decoder \\ block};
\node (dec3) [rect2] at (3.5, 8.3) {$(D)^{\textrm{th}}$ Decoder \\ block};
\node (sf3) [emb]    at (3.5, 9.9) {$\mathrm{softmax}$ \\ layer};
\node (pd3) [var]    at (3.5, 11.3) {$\mathcal{L}_{D}^{\text{Attn.}}$};

\node (ca)  [var]    at (1.2, 2.6) {C/A};
\node (sn) [hidden]  at (-0.1, 2.3)   {};
\node (a0)  [hidden] at (1.7, 2.3) {};
\node (a1)  [hidden] at (1.7, 4.3) {};
\node (a2)  [hidden] at (1.7, 6.3) {};
\node (a3)  [hidden] at (1.7, 8.3) {};

\node (i0)  [hidden] at (5.3, 5.3) {};
\draw [arr] (x)   edge (enc);
\draw [arr] (enc) edge (ctc);
\draw [arr] (ctc) edge (pctc);
\draw [arr] (sf1) edge (pd1);
\draw [arr] (sf3) edge (pd3);

\draw (sn) edge (a0);
\draw (a0) edge (a1);
\draw (a1) edge (a2);
\draw (a2) edge (a3);
\draw (h13) edge (i0);
\draw [arr] (a0) edge (dec0);
\draw [arr] (dec1) edge (dec2);
\draw [arr] (a1) edge (dec1);
\draw [arr] (a2) edge (dec2);
\draw [arr] (dec2) edge (dec3);
\draw [arr] (a3) edge (dec3);
\draw [arr] (dec3) edge (sf3);

\draw [arr] (i0) edge (sf1);
\end{tikzpicture}
\caption{Architecture of the proposed DeCRED.
In addition to the standard encoder-decoder framework for ASR ($\mathcal{L}_D^{\text{Attn}}$), with the auxiliary CTC objective ($\mathcal{L}^\text{CTC}$), DeCRED uses -- possibly multiple -- auxiliary classifiers ($\mathcal{L}_d^{\text{Attn}}$) attached to the decoder. In the illustration, we show one auxiliary classifier attached to $(D\text{-}2)$-th decoder block. The embedding and positional encoding layers are not depicted for brevity.}
\label{fig:framework_architecture}
\end{figure}

\section{Decoder-centric regularization}
\label{sec:decred}
Formally, our approach extends the training objective of encoder-decoder ASR by adding auxiliary cross-entropy loss functions. We explore two additional decoding methods that exploit these auxiliary classifiers.

\subsection{Training objective}
\label{sec:objective}
We build upon the hybrid CTC-attention-based training scheme proposed by~\citet{hori_joint_2017}. Our objective function $\mathcal{L}$ is defined as:

\begin{equation}
\mathcal{L} = \alpha \,\mathcal{L}^{\text{CTC}} + (1 - \alpha)\, \mathcal{L}^{\text{DeCRED}},
\end{equation}
where $\mathcal{L}_{\text{CTC}}$ represents the standard CTC loss~\cite{graves_connectionist_2006}, $\alpha$ is a hyper-parameter, and $\mathcal{L}_{\text{DeCRED}}$ is defined as:
\begin{equation}\label{eq:decred}
\mathcal{L}^{\text{DeCRED}} = \sum_{d=1}^{D} \beta_{d} \mathcal{L}^{\text{Attn}}_{d},
\end{equation}
where $D$ represents the number of layers in the decoder, $\mathcal{L}^{\text{Attn}}_{d}$ is the cross-entropy loss given a classifier layer (linear projection, followed by softmax function) attached to the $d$-th layer of the decoder, and $\beta_{d}$ is the weighting factor of $d$-th layer. We impose constraints such that $\sum_{d=1}^{D} \beta_{d} = 1$ and $\beta_{d} \geq 0$. In practise $[\beta_1 \ldots \beta_D]$ is a sparse vector.
This definition allows us to explicitly regularise the decoder (internal language model) and force earlier layers to learn discriminative features suitable for the task.
Figure~\ref{fig:framework_architecture} illustrates the proposed architecture, where an auxiliary classifier is attached to the output of $(D\text{-}2)$-th decoder block.

\subsection{Decoding}
\label{sec:method_decoding}
The decoding follows a typical auto-regressive scheme observed in encoder-decoder ASR systems, where the posterior probability of an output token is obtained by conditioning on previously decoded tokens (partial hypothesis) and the input features.

Formally, let $\mathbf{x}_{1:T}$ be a sequence of input speech (filterbank) features and 
let $y_{1:N}$ be a sequence of output tokens.
Following the joint CTC/attention decoding~\cite{hori_joint_2017}, the posterior probability of an output token $y_n$ is evaluated as

\begin{align}
 \label{eq:decoding}
 {}& \log p(y_n \mid y_{1:n-1}, \mathbf{x}_{1:T}) \approx \nonumber\\ 
  & \,\, \lambda \, \log p_{\text{CTC}}(y_n\mid y_{1:n-1},\mathbf{x}_{1:T}) \, + \, \nonumber \\
 &\,\, (1 -\lambda) \, \log p_{\text{DeCRED}}(y_n\mid y_{1:n-1},\mathbf{x}_{1:T}), 
 \end{align}
where $\lambda$ is a hyper-parameter. 

Now, let $\mathbf{h}_{d,n} \in \mathbb{R}^{1 \times d_{\text{model}}}$ denote the hidden representation corresponding to the $n$-th output token obtained from the $d$-th layer of the decoder, and $\mathbf{W}_{d} \in \mathbb{R}^{d_{\text{model}} \times V}$ represent linear projection from hidden dimension $d_{\text{model}}$ to vocabulary size $V$. 
We obtain the following decoding methods by varying the definition of $p_{\text{DeCRED}}$:

\begingroup huggi
\begin{enumerate}
    \item Vanilla joint CTC/attention decoding relying on representations \textit{only} from the last layer:
    \begin{flalign}\label{eq:decoding-baseline}
         p_{\text{DeCRED}}(y_n\mid y_{1:n-1},\mathbf{x}_{1:T}) = \nonumber\\ 
         \text{softmax}(\mathbf{h}_{D,n} \mathbf{W}_{D})
    \end{flalign}
    \item Sum of logits weighted by per-layer learnable scalar $\beta_{d}$:
    \begin{flalign}\label{eq:decoding-per-layer}
     p_{\text{DeCRED}}(\cdot) =  \text{softmax}(\sum_{d=1}^{D} \beta_{d} \mathbf{h}_{d,n} \mathbf{W}_{d})
    \end{flalign}
    \item  Sum of logits weighted by  per-layer learnable vector $\mathbf{v}_{d} \in \mathbb{R}^{1 \times V}$, where $\odot$ is elementwise product:
        \begin{flalign}\label{eq:decoding-per-token}
             p_{\text{DeCRED}}(\cdot) =  \text{softmax}\left (\sum_{d=1}^{D} \mathbf{v}_{d} \odot (\mathbf{h}_{d,n} \mathbf{W}_{d}) \right)
        \end{flalign}
\end{enumerate}
\endgroup
Note that to obtain optimal results with methods~\eqref{eq:decoding-per-layer} and~\eqref{eq:decoding-per-token}, an additional held-out set is required for learning the parameters $\beta_d$, $\mathbf{v}_d$. 

The above schemes can be easily integrated into any of the decoding search algorithms, such as greedy and beam-search.

\begin{figure*}[t]
    \centering
    \includegraphics[width=\linewidth]{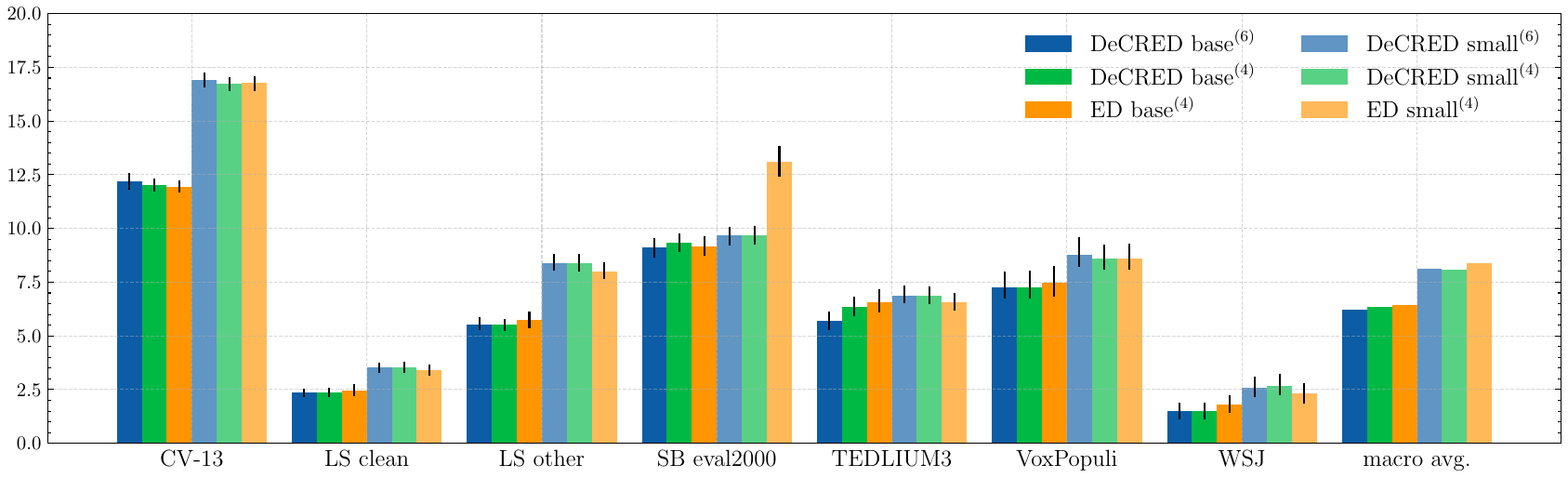}
    \caption{The impact of employment of the proposed training strategy, along with the enhanced decoding~\eqref{eq:decoding-per-token} on small (12, 6, 256, 5000) and base (16, 8, 512, 5000)  models using greedy $\lambda =0.3$ decoding. $\text{DeCRED-base}^{(\ref{eq:decoding-per-token})}$ indicates the model with the proposed decoding technique~\eqref{eq:decoding-per-token}, and the mixing parameters $\mathbf{v}$ tuned on development split. To compute confidence intervals, we employed bootstrapping with $\alpha=0.05$ and $B=1000$.}
    \label{fig:ed_vs_decred}
\end{figure*}

\begin{figure*}[t]
    \centering
    \includegraphics[width=\linewidth]{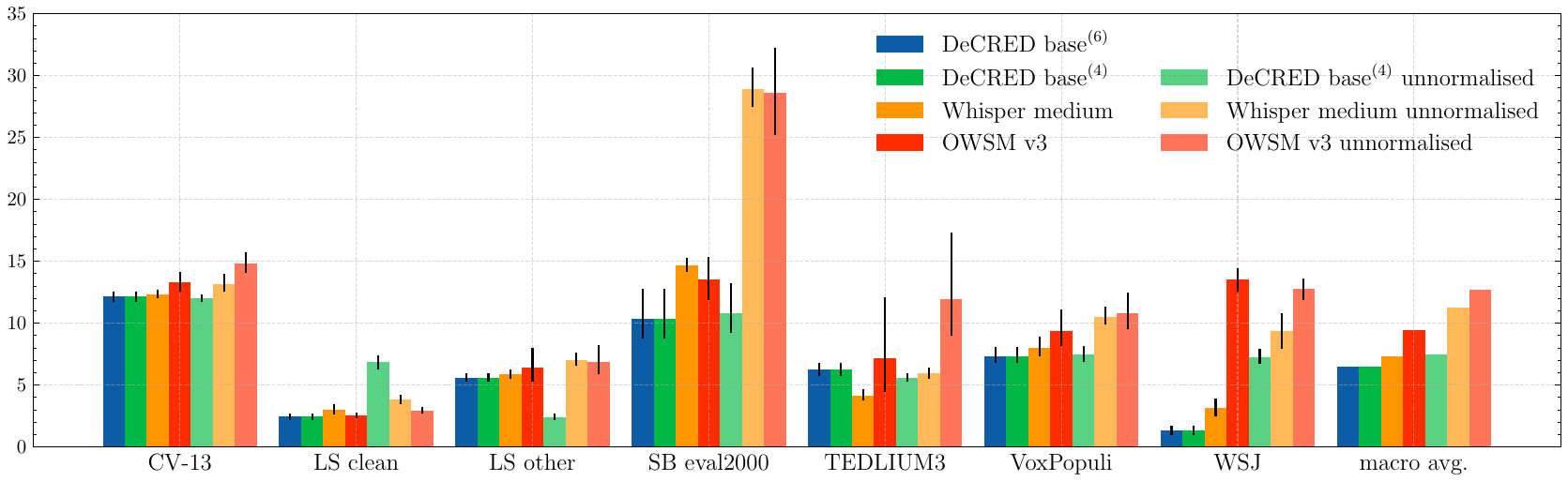}

    \caption{Comparison of the proposed model against publicly available models on original and normalised transcripts using greedy decoding~ \eqref{eq:decoding-baseline} with $\lambda = 0$, as Whisper lacks a CTC head. Additional gains can be observed when using $\lambda = 0.3$ for both DeCRED and OWSM v3 models.}
    \label{fig:whisper_comparison}
\end{figure*}

\section{Experimental setup}
\label{sec:exp}
The experiments are organised into two parts.
The first part (Sec.~\ref{sec:multi-domain}) compares baseline ED with the proposed DeCRED on multi-domain English datasets. The selection of datasets is inspired by those used for evaluation in OWSM. This relatively larger corpus allows us to fully exploit the proposed decoding alternatives~\eqref{eq:decoding-per-layer} and~\eqref{eq:decoding-per-token}, assess the effectiveness of Internal Language Model (ILM) regularisations, and compare our models'  in- and out-of-domain performance with large-scale trained speech models.

The second part (Sec.~\ref{sec:ablation}) focuses on single in-domain datasets,
studying the effects of the position ($d$), weight ($\beta_d$) of the auxiliary classifiers, their influence in decoding and the impact of single-domain text normalisations.

All our experiments are built on the open-source \texttt{transformers} library, accompanied by baseline models built using the ESPnet toolkit.

\subsection{Baseline Encoder-Decoder (ED) model}
\label{sec:baseline}
Throughout this paper, we use the quadruplet ($E$, $D$, $d_{\text{model}}$, $V$), where $E$ represents the number of layers in the encoder, $D$ refers to the decoder layers, $d_{\text{model}}$ is the hidden dimension, and $V$ is the vocabulary size. The rest of the configuration remains fixed unless explicitly stated otherwise.

Our baseline ED small $(12, 6, 256, 5000)$ and ED base $(16, 8, 512, 5000)$ models contain 39M and 172M parameters, respectively. They receive 80-dimensional filter-bank features as input and employ an input module consisting of two Conv2d layers with 256 output channels, followed by a linear projection. This is followed by an $E$-layer E-Branchformer~\cite{kim_e-branchformer_2023} encoder with relative positional embeddings~\cite{dai_transformer-xl_2019}, Macaron-like feedforward modules~\cite{gulati_conformer_2020},  $d_{\text{ff}} = 4d_{\text{model}}$, four attention heads, and a dropout probability of 0.1.

In line with the E-Branchformer architecture, we incorporate a merge block followed by depth-wise convolution with a kernel size of 31. The encoder is followed by a $D$-layer Transformer decoder with sinusoidal positional embeddings, maintaining the same number of attention heads, $d_{\text{model}}$, and dropout. The  decoder has fixed $d_{\text{ff}} = 2048$. We use a subword tokenizer based on the unigram algorithm.

We use the same quadruplet as in ED to define the Decoder-Centric Regularized Encoder-Decoder (DeCRED) architecture. The only difference is the number of attached classifiers and their corresponding weights. The weights of the additional classifiers are not tied to the one attached to the $D$-th layer. For the baseline DeCRED models (small and base), a single additional classifier with $\beta_{D-2} = 0.4$ is attached, adding only $d_{\text{model}} \times V$ new parameters.

\subsection{Training details}
\label{sec:training_details}
Our models are trained on Nvidia A100 GPUs with bf16 precision using the AdamW optimiser~\cite{loshchilov2018decoupled} for $100$ epochs with early stopping patience of $10$, learning rate of $2\times 10^{-3}$, weight decay of $1\times 10^{-6}$, linear decay scheduler and 40k warm-up steps. We use a label smoothing~\cite{PereyraTCKH17, watanabe_espnet_2018} weight of $0.1$ as an additional means of regularisation. To speed up the training, samples longer than 20 seconds are discarded from the training set. 

Unlike ESPnet, where some augmentations are applied offline, we implement all augmentations online and allow for postponing some of them until later in the training, resulting in a more stable training process.
For instance, while ESPnet adopts a training regime consisting of 50 epochs and three copies of the input data with speed perturbation factors 0.9, 1.0, and 1.1, we train our model for 150 epochs on the original data with speed perturbation factors $\{0.9, 1.0, 1.1\}$ randomly sampled on the fly.
After 5k update steps, we apply SpecAug~\cite{park_specaugment_2019} with two frequency-masks of maximum size $27$ and five time-masks with maximum coverage of masked input of $5\,\%$.
For all experiments, we select the best-performing checkpoint based on the development WER.
Additionally, we introduce a mechanism to mask special tokens, along with unfinished words\footnote{e.g.\ transcript ``[hesitation] to re- to re- renew" is transformed into ``[MASK] to [MASK] to [MASK] renew"}, during error backpropagation.
This strategy aimed to prevent the model from being penalised for unclear inputs.

\section{ED vs DeCRED in multi-domain scenario}
\label{sec:multi-domain}
To build robust ASR systems that are on par with state-of-the-art, we chose a mixture of multi-domain datasets that allows for bigger training, development and test sets. 
The multi-domain dataset is comprised of Fisher (SWITCHBOARD)~\cite{godfrey_switchboard_1992},  WSJ~\cite{paul_design_1992}, Common Voice en 13~\cite{ardila_common_2020}, LibriSpeech~\cite{panayotov_librispeech_2015}, VoxPopuli~\cite{wang_voxpopuli_2021}, and TED-LIUM~3~\cite{hernandez_ted-lium_2018}, totalling 6k hours of training data. To study the generalisation capabilities of our models, we also evaluate our ED and DeCRED models on three unseen datasets (AMI~\cite{AMI}, FLEURS~\cite{conneau_fleurs_2023} and Gigaspeech~\cite{gigaspeech}).

\begingroup
\setlength{\tabcolsep}{5pt} 
\begin{table}[t]
    \centering
    \caption{%
        Comparison of ED and DeCRED models on out-of-domain test sets.
        WERs are obtained using greedy decoding with $\lambda=0$.
        $\dagger$ denotes models where $\mathbf{v}^{*}$ was tuned on each of the datasets separately. 
    }\label{tab:ood}
    \small{\begin{tabular}{lccc}
        \toprule
        \text{Model} & FLEURS & AMI-ihm & Gigaspeech \\
        \midrule
        $\text{ED base}^{(\ref{eq:decoding-baseline})}$ & 6.4  & 24.8  & 19.8  \\
        $\text{DeCRED base}^{(\ref{eq:decoding-baseline})}$ & 6.7  & 22.1  & 16.9  \\
        $\text{DeCRED base}^{(\ref{eq:decoding-per-token})}$ & 6.9 & 21.9 & 17.0  \\
        $\text{DeCRED base}^{(\ref{eq:decoding-per-token})\dagger}$ & 6.8  & 21.4 & 16.4 \\
        \midrule
        OWSM v3 & 8.6 & 35.8 & 34.1   \\
        Whisper medium & \textbf{5.5} & \textbf{16.6} & \textbf{14.9 }\\
        
        \bottomrule
    \end{tabular}}
\end{table}
\endgroup

\subsection{Normalisation of multi-domain data}
These datasets have different annotation styles, making learning harder and introducing undesired behaviour in the models, such as memorising the dataset-specific annotations~\cite{peng_reproducing_2023}.
We employed a practical approach using the text normalisation scheme from Whisper\footnote{\url{https://github.com/huggingface/transformers/blob/main/src/transformers/models/whisper/english_normalizer.py}} to standardise the transcripts across all the datasets.
We believe this approach allows the model to focus mainly on the recognition task.
For practical applications, true casing and punctuation can be restored using a lightweight inverse text normalisation model. In addition to the Whisper text normaliser, we retained the text within parenthesis.
Due to inconsistencies across datasets, we removed special tokens such as [breath], [vocalised noise], [pause], [sneeze].

Nevertheless, to enable a fair comparison with prior works, we also report results by training and evaluating the original transcripts.
This allows us to quantify the effect of text normalisation on WER.

\begingroup
\setlength{\tabcolsep}{3pt} 
\begin{table*}
    \centering
    \caption{%
        Zero-Attention Internal Language Model (ILM) BPE-level perplexity estimation of ED and DeCRED models on in- and out-of-domain test sets.
    }\label{tab:ppl}
    \small{\begin{tabular}{l|ccccccc|ccc}
        \toprule
        \text{Model}  & CV-13 & LS clean & LS other & SB eval2000 & TEDLIUM3 & VoxPopuli & WSJ & FLEURS & AMI-ihm & Gigaspeech  \\
        \midrule
        $\text{DeCRED}$ & 141.0 & 129.1 & 140.4 & 104.1 & 89.0   & 101.4 & 126.0  & 111.5 & 136.6 & 66.3 \\
        $\text{ED}$     & 232.4 & 206.1 & 199.9 & 220.3 & 134.6  & 142.7 & 177.3  & 159.7 & 308.3 & 84.0 \\
        \bottomrule
    \end{tabular}}
\end{table*}
\endgroup

\subsection{Comparison with a fair baseline}\label{sec:decred_vs_ed}
Figure~\ref{fig:ed_vs_decred} compares the WER of baseline ED and the proposed DeCRED across all the in-domain datasets. Specifically, we compare both the small (12, 6, 256, 5000) and base (16, 8, 512, 5000) variants of both ED and DeCRED.

To learn the mixing parameters $\beta_{d}^{}$ and $\mathbf{v}^{}$ for the respective decoding methods (Section~\ref{sec:method_decoding}), we followed the Platt scaling approach~\cite{Guo_calibrating_2017, Lee_calibration_2021}, splitting the original set development utterances into new training and development sets with a 70:30 ratio. Except for the mixing parameters, the rest of the model remains frozen. This training, or fine-tuning, is very lightweight.

We use the equation number in the superscript of the model to denote the decoding objective, i.e. DeCRED\textsuperscript{\eqref{eq:decoding-baseline}} indicates the vanilla decoding method defined by~\eqref{eq:decoding-baseline}, DeCRED\textsuperscript{\eqref{eq:decoding-per-layer}} indicates mixing the logits by learnable scalars, and DeCRED\textsuperscript{\eqref{eq:decoding-per-token}} indicates mixing the logits by learnable vectors. 

The Figure also shows the macro WER computed across all the datasets. We computed a statistical significance test using the bootstrapping method, which showed that results from $\text{DeCRED-base}^{(\ref{eq:decoding-baseline})}$ and $\text{DeCRED-base}^{(\ref{eq:decoding-per-token})}$ are statistically significant than baseline $\text{ED-base}^{(\ref{eq:decoding-baseline})}$
 with $p$ values of $0.35$ and $0.3$ respectively.

\begin{figure}[t]
    \centering
    \includegraphics[width=\linewidth]{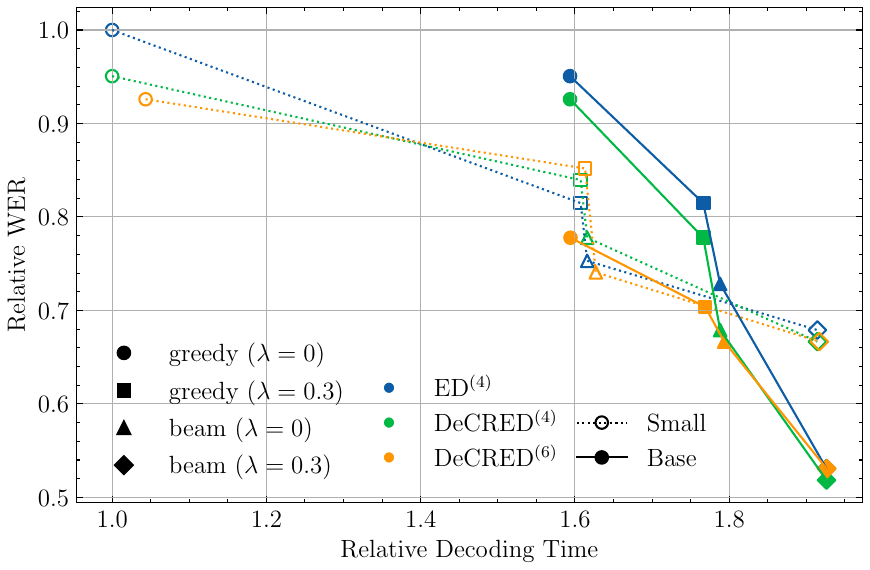}
    \caption{%
        The impact of model size and decoding approach on the average time needed to transcribe an utterance (TEDLIUM3) and WER (macro average across datasets).    
    }\label{fig:slowdown_wer}
\end{figure}

\subsection{Comparison with Whisper and OWSM}
Figure~\ref{fig:whisper_comparison} provides a reference comparison between our best models, $\text{DeCRED-base}^{(\ref{eq:decoding-baseline})}$ (172M parameters) and $\text{DeCRED-base}^{(\ref{eq:decoding-per-token})}$, and large-scale multilingual models — Whisper-medium~\cite{radford_robust_2023} (700M parameters) and OWSM v3~\cite{peng_reproducing_2023} (889M parameters).

It is important to note that this figure serves only as a reference and does not represent a fully fair comparison, as the models differ significantly in terms of scale and design.

Although most of our models were trained on normalised transcriptions, we also trained the DeCRED-base (16, 8, 512) model on original transcriptions to highlight the effect of text normalisation.
To ensure consistency, we applied the same text normalisation used in our training pipeline to the Whisper and OWSM outputs during evaluation in the normalised setup.

\subsection{Performance on out of domain }\label{sec:out_domain}
In Table~\ref{tab:ood}, we compare the performance of ED and DeCRED models on out-of-domain datasets that were not seen during training. We take this as an opportunity to evaluate the effect of rapid tuning of the mixing weights $\mathbf{v}^{*}$ on the corresponding domain.
For this, we utilised FLEURS train split and development splits of AMI and Gigaspeech, respectively, following the training protocol described in Section~\ref{sec:method_decoding}.
In Table~\ref{tab:ood}, these models are denoted as $\text{DeCRED base}^{(\ref{eq:decoding-per-token})\dagger}$.
In all cases, adapting $\mathbf{v}^{*}$ leads to a decrease in WER, and with the exception of the FLEURS dataset, this decrease is considerable, confirming that the mixing weight does provide a rapid adaptation capability. 

Overall, all our models outperform the much larger OWSMv3, which has also been trained on the corresponding training data, showing that our models do generalize to unseen domains well.
With the exception of the FLEURS dataset, where the difference in WER is the smallest anyway, DeCRED models outperform the ED baseline significantly, suggesting that the decoder-centric regularisation enhances the model's generalisation ability.

\begingroup
\setlength{\tabcolsep}{3pt} 
\begin{table}[t]
    \centering    
    \caption{%
        Macro average of the WERs based on the selected decoding. All auxiliary classifier weights $\beta_{d}$ are set to $0$ by default. Parameters with an asterisk (e.g., $\beta_{d}^{*}$, $\mathbf{v}^{*}$) indicate tuning on a portion of the development split, i.e. training mixing weights for layers $D-2$ and $D$ for DeCRED-base.
    }\label{tab:decoding_analysis}
    \small{
    \begin{tabular}{ccccc}
    \toprule
    
    DeCRED-base                     & \multicolumn{2}{c}{greedy} & \multicolumn{2}{c}{beam -- width 10}       \\
    decoding strategy                          & $\lambda=0$   & $\lambda=0.3$ & $\lambda=0$   &  $\lambda=0.3$    \\
    \midrule
    $\beta_{6} =1^{(\ref{eq:decoding-per-layer})}$                              & 6.8    & 6.7  & 6.5   & 6.4\\
    $\beta_{8} =1^{(\ref{eq:decoding-baseline})}$                              & 6.7   & 6.4   & \textbf{6.4} & \textbf{6.0} \\
    $\beta_{6,8}=(0.40, 0.60)^{(\ref{eq:decoding-per-layer})}$           & 6.7  & 6.6  & \textbf{6.4} & 6.3 \\
    $\beta_{6,8}^{*}=(0.50, 0.47)^{(\ref{eq:decoding-per-layer})}$   & 6.7  & 6.6  & 6.5 & 6.3\\
    $\mathbf{v}^{*(\ref{eq:decoding-per-token})}$                            & \textbf{6.5}   & \textbf{6.3}   & 6.9   & \textbf{6.0}             \\
    \bottomrule
    \end{tabular}
    }
\end{table}
\endgroup

\subsection{Comparison across different decoding methods}
Table~\ref{tab:decoding_analysis} presents a comparison of different decoding methods in terms of macro WER over the in-domain datasets.
We observed that integrating intermediate representations with per-layer learnable weights~\eqref{eq:decoding-per-layer} led to minor improvement only in the greedy decoding scenario without hybrid CTC decoding.
Notable improvements were observed with the incorporation of per-token-specific mixing~\eqref{eq:decoding-per-token}, except for beam decoding on the SB eval2000 dataset, where we observed a degradation of $5.3\,\%$ in WER, primarily attributed to insertions.
Interestingly, we did not observe the same behaviour with the small model.
For completeness, macro WERs are also provided for early-exiting ($\beta_6=1$, i.e., decoding directly from the 6th layer, while the model has 8 layers), where only minor degradations were observed.

\subsection{Trade-off between performance and decoding time}

Figure~\ref{fig:slowdown_wer} presents the relative WER reductions of our multi-domain models on TEDLIUM3 in relation to the relative slowdown caused by additional decoding overhead. The slowdown factor is measured relatively to our fastest model $\text{ED small}^{(\ref{eq:decoding-baseline})}$. It is calculated as an average time over the TEDLIUM3 test set required to emit 20 tokens on an A100 GPU with maximum VRAM memory consumption\footnote{For example, with greedy decoding and ED small, we can fit 240 samples in a batch. In contrast, with ED base and joint CTC/attention decoding with a beam size of 10, we are only able to fit 20 samples.}. We fixed a number of decoding steps to normalise different hypothesis lengths across models. 

In this setup, there is no speed difference between $\text{ED}^{(\ref{eq:decoding-baseline})}$ and $\text{DeCRED}^{(\ref{eq:decoding-baseline})}$. However, as shown in Figure~\ref{fig:slowdown_wer}, regularised models significantly reduce the WER. When using $\text{DeCRED}^{(\ref{eq:decoding-per-token})}$, the only overhead is computing $\text{softmax}\left (\sum_{d=1}^{D} \mathbf{v}_{d} \odot (\mathbf{h}_{d} \mathbf{W}_{d}) \right)$, where $\mathbf{h}_{D} \mathbf{W}_{D}$ is already computed.
It is worth noting that when using greedy decoding, DeCRED small performs similarly to ED base, being much smaller, thus consuming less computation resources and speeding up decoding significantly.

\begin{table}[t]
\centering
\caption{Comparison of our implementation of ED and proposed DeCRED with InterCTC and ESPnet's ED baselines on the TEDLIUM3 test split.
}\label{tab:espnet_v_hf}
\small{
\begin{tabular}{lp{1.1cm}cc}
\toprule
        &                       & \multicolumn{2}{c}{WER [\%]} \\
\text{Model} & Size [M] & greedy & beam -- width 40 \\
\midrule
ESPnet $\text{ED}^{(\ref{eq:decoding-baseline})}$ & 35.01 & 8.7 & 8.1 \\
Our $\text{ED}^{(\ref{eq:decoding-baseline})}$ & 35.04 & 7.6 &7.2 \\
$\text{DeCRED}^{(\ref{eq:decoding-baseline})}$ & 35.20 & \textbf{7.0} & \textbf{6.8} \\

$\text{InterCTC}^{(\ref{eq:decoding-baseline})}_{\lfloor L/2 \rfloor}$ & 35.20 & 7.5 & 7.1 \\
\bottomrule
\end{tabular}
}
\end{table}

\section{Analysis of internal language model}\label{sec:ilm}
In the attention-based Encoder-Decoder (ED) ASR framework, the decoder functions as an autoregressive internal language model. In this paper, we directly regularise this component of the network by incorporating auxiliary classifiers. In the previous Section~\ref{sec:multi-domain}, we demonstrated consistent improvements of DeCRED vs ED achieved through such regularisation. Building on the work of \citet{zeineldeen_investigating_2021}, we employ Zero-Attention Internal Language Model (ILM) subword-level perplexity estimation to analyse the impact of our proposed regularisation scheme across in-domain and out-of-domain datasets.

Table~\ref{tab:ppl} showcases a consistent reduction in perplexity across all analysed datasets of DeCRED base versus ED base, strongly indicating that the ILM generalises much better across multiple domains, which further supports our hypothesis.

\section{Ablations on in-domain dataset}
\label{sec:ablation}
To further analyse and understand the proposed regularisation scheme, we select a relatively small dataset, TEDLIUM3~\cite{hernandez_ted-lium_2018}, which allows for faster experiment turnout. The dataset comprises 452 hours of transcribed TED talks, with a test set containing 1155 utterances, roughly translating to 28k words.
The size of this dataset enables us to train an ED (12, 6, 256, 500) baseline 35M model to full convergence in approximately 70 A100 hours.

Since we build on top of \texttt{transformers} library, to ensure a fair comparison, we adopt hyperparameters and a training setup as close as possible to the ESPnet baseline recipe\footnote{\url{https://github.com/espnet/espnet/tree/master/egs2/tedlium3/asr1}}.
For evaluating the models on TEDLIUM3, unless explicitly specified, we follow the ESPnet recipe utilising joint CTC/attention decoding~\eqref{eq:decoding-baseline} with a beam size of 40 and CTC decoding weight $\lambda=0.3$.

Table~\ref{tab:espnet_v_hf} compares our best-performing DeCRED and ED baseline models with the baseline model from ESPnet and InterCTC baseline~\cite{lee_intermediate_2021}.
DeCRED consistently outperforms both implementations of ED.
The difference is better pronounced in greedy decoding, suggesting the effectiveness of DeCRED in decoding tasks where computational resources are limited.
Appendix~\ref{ssec:arch_search} provides more details about the hyperparameter search.

\subsection{Effect of text normalisation}
To further understand the effect of text normalisation, we trained standalone models ED-small (12, 6, 256) on the TEDLIUM3 and Voxpopuli datasets with and without normalised transcripts. Notably, with normalisation, we observed an improvement from $9.8\,\%$ to $9.0\,\%$ WER on VoxPopuli and from $7.2\,\%$ to $6.7\,\%$ on TEDLIUM3. The normalisation process effectively resolved contraction errors and also led to fewer errors in the most frequent confusion pairs (e.g., ``the'' vs ``a'', ``in'' vs ``on'', ``in'' vs ``and''). By normalising, we reduced the number of words from 44.3k to 44.1k for Voxpopuli and increased this number from 27.5k to 28.2k for TEDLIUM3, which also influenced the WER.

\section{Conclusion}
We introduced the DeCRED regularization scheme, which effectively integrates auxiliary classifiers within the decoder of an encoder-decoder-based architecture.
We further proposed decoding methods that exploit these auxiliary classifiers, which led to a significant decrease in the word error rates.
We observed that DeCRED consistently improves the results when employing a simple greedy decoding scheme compared to the baseline models.
Our experiments on multi-domain datasets show that DeCRED is scalable, performs competitively to much larger Whisper medium, and outperforms OWSM v3.
Finally, we show that DeCRED enhances the generalisation to out-of-domain datasets, where we observed a reduction of 2.7\,\% and 2.9\,\% WER, on AMI and Gigaspeech, respectively.
Using a lightweight rapid domain adaptation scheme enabled by DeCRED, the out-of-domain WERs were further reduced by 0.7\,\% and 0.5\,\% absolute on the respective datasets.
In future, we intend to study DeCRED in multilingual and multi-task scenarios.

\section{Limitations}
We identify a few limitations in our work.
Firstly, due to our computational budget, we were only able to scale our setup to 6k hours of training data and 172M model parameters.
Secondly, our models were trained on English data only, which makes the comparison with multilingual models tricky, as these models had to invest a part of their capacity into modelling other languages as well.
Yet, due to the first point, our models are exposed to one (OSWM) or even two (Whisper) orders of magnitude less English data, therefore we believe the comparison is not unfair.
Also, our models use considerably smaller vocabulary; however, while this might limit model performance on domain-specific words present, for example, in the FLEURS dataset, we do not observe performance degradation there.

Next, some of the improvements from introducing DeCRED diminish when employing beam-search decoding with a wider beam, which, however, comes at a computational cost at inference time.
Finally, while the proposed decoder-centric regularization is independent of the backbone architecture, we have only analysed our approach using an E-branchformer speech encoder.

\bibliography{custom}

\appendix

\section{Position and weight of auxiliary classifiers}
\label{ssec:arch_search}
Even with a model with as few as $D=6$ decoder layers, the definition of the DeCRED objective~\eqref{eq:decred} leaves us with a vast configuration space.
We explored this space starting with the configurations with a single auxiliary classifier, changing its position and adjusting its weight in increments of $0.1$.
The additional parameters introduced ($\textbf{W}_{d} \in \mathbb{R}^{256 \times 500}$) by a single auxiliary classifier do not significantly increase the model size.

The results are summarised in Table~\ref{tab:location_weight}. 
Compared to our baseline ED model with a WER of 7.2\,\%, we observe improvements with the additional classifier placed closer to the final layer.

Further experiments with multiple auxiliary classifiers ($\{\beta_3=0.2, \beta_4=0.3, \beta_6=0.5\}$ and $\{\beta_3=0.2, \beta_5=0.3, \beta_6=0.5\}$), did not yield significant improvements, discouraging experiments with more auxiliary classifiers. 
We avoided exploring very low weights $(\beta_d)$ in the early layers as gradual adjustments did not yield noticeable improvements.
Given the computational resources required for each experiment run, we chose to run the two most promising configurations five times to determine the optimal one.
Choosing between the two most promising configurations from Table~\ref{tab:location_weight}, i.\thinspace e., $\beta_{3} = 0.5$ and $\beta_{4} = 0.4$, we opted for the latter for all subsequent experiments. We believe other configurations (indicated with grey colour in the lower triangle of Table~\ref{tab:location_weight}) could yield similar results.

\begingroup
\setlength{\tabcolsep}{4pt} 
\begin{table}[!ht]
    \centering
    \caption{%
        Effect of the position ($d$) and weight ($\beta_d$) of the auxiliary classifier in DeCRED on WERs of TEDLIUM3 test set. Grey cells indicate configurations deemed reasonable for exploration.
        Standard deviations ($\sigma$) and best WER for the chosen configurations are displayed. For reference, the baseline ED model has a WER of 7.2\,\%.
    }\label{tab:location_weight}
    \small{
    \begin{tabular}{clllll}
    \toprule
    \multirow{2}{*}{\text{Weight}} & \multicolumn{5}{c}{\text{Position}}\\
    \cmidrule(l){2-6}
     & 1     & 2     & 3     & 4     & 5     \\
    \midrule
    0.1 &       &       &               & 7.5  & 6.8 \\
    0.2 &       &       &               & 7.0  & 7.2 \\
    0.3 &       &       & 7.0          & \cellcolor{gray!25}7.0  & \cellcolor{gray!25}7.0 \\ 
    0.4 & 7.8  & 7.5  & \cellcolor{gray!25}7.1          & \cellcolor{gray!25}\textbf{6.8} [$\sigma = 0.15$]  & \cellcolor{gray!25}6.9 \\
    0.5 & 7.2   & 7.1  & \cellcolor{gray!25}\textbf{6.7} [$\sigma = 0.26$] & \cellcolor{gray!25}7.1  & \cellcolor{gray!25}6.9 \\ 
    \bottomrule
    \end{tabular}
    }
\end{table}

\endgroup

\end{document}